\documentclass[particles,conferenceproceedings,submit,moreauthors,pdftex]{mdpi} 

\usepackage{subfigure}
\usepackage{dsfont}
\usepackage{slashed}

\newcommand{\Ml}{\mathcal{M}_{ud}}
\newcommand{\Ms}{\mathcal{M}_{s}}
\newcommand{\ml}{m_{ud}}
\newcommand{\pion}{{\pi^\pm}}
\newcommand{\Z}{\mathcal{Z}}
\newcommand{\D}{\mathcal{D}}
\newcommand{\Dn}{\slashed{D}(0)}
\newcommand{\Dp}{\slashed{D}(\mu_I)}
\newcommand{\Dm}{\slashed{D}(-\mu_I)}
\firstpage{1} 
\pubvolume{xx}
\issuenum{1}
\articlenumber{5}
\pubyear{2019}
\copyrightyear{2019}
\history{Received: date; Accepted: date; Published: date}
\definecolor{bluecite}{HTML}{0875b7}
\usepackage[colorlinks=true, 
	linkcolor=bluecite, 
	citecolor=bluecite]{hyperref}

\Title{Dirac spectrum and the BEC-BCS crossover in QCD at nonzero isospin asymmetry}

\Author{B. B. Brandt, F. Cuteri, G. Endr\H{o}di and S. Schmalzbauer}
\AuthorNames{B. B. Brandt, F. Cuteri, G. Endr\H{o}di and S. Schmalzbauer}

\address[1]{%
Institute for Theoretical Physics, Goethe University \\Max-von-Laue-Strasse 1, 60438 Frankfurt am  Main, Germany\\
E-mail:  \{brandt,cuteri,endrodi,schmalzbauer\}@itp.uni-frankfurt.de}

\abstract{
For large isospin asymmetries, perturbation theory predicts the QCD ground state to be a superfluid phase of $u$ and $\bar{d}$ Cooper pairs.
This phase, which is denoted as the BCS phase, is expected to be smoothly connected to the standard phase with Bose-Einstein condensation (BEC) of charged pions at $\mu_I\ge m_\pi/2$ by an analytic crossover.
A first hint for the existence of the BCS phase, which is likely characterised by the presence of both, deconfinement and charged pion condensation, is coming from the lattice observation that the deconfinement crossover smoothly penetrates into the BEC phase.
To further scrutinize the existence of the BCS phase, in this proceedings article we investigate the complex spectrum of the massive Dirac operator in 2+1-flavor QCD at nonzero temperature and
isospin chemical potential. The spectral density near the origin is related to the BCS gap via a generalization of the Banks-Casher relation to the case of complex Dirac eigenvalues (derived for the zero-temperature, high-density limits of QCD at nonzero isospin chemical potential).
}


\begin{document}
%%%%%%%%%%%%%%%%%%%%%%%%%%%%%%%%%%%%%%%%%%

%%%%%%%%%%%%%%%%%%%%%%%%%%%%%%%%%%%%%%%%%%

\section{Introduction}
Quantum chromodynamics (QCD) is established as the fundamental theory governing nuclear matter and hadrons.
It holds that hadrons are made from quarks, their antimatter siblings, and gluons that carry the strong/color force binding quarks to each other while being themselves color charged objects.
The fundamental laws of QCD are elegantly concise, however, understanding the structural complexity of hadrons in terms of quarks and gluons governed by those laws remains an open challenge.

In the highest energy RHIC and LHC collisions, strongly interacting matter at the relevant energies contains almost as many antiquarks as quarks, which, borrowing condensed matter physics nomenclature, one could call ``undoped strongly interacting matter''.
However, there exist many physical settings, like non-central heavy-ion collisions, the structure of compact stars and the evolution of the early Universe, where instead strongly interacting matter is doped with e.g. an excess of down quarks over up quarks. In the physical systems mentioned above this translates into an excess of neutrons over protons or positively charged pions over negatively charged pions. 
To understand these systems, we must map the phase diagram of QCD as a function of both temperature and ``doping'', which we can express in terms of (negative) isospin density $n_I=n_u-n_d$ or, equivalently, in the grand canonical approach to QCD, in terms of isospin chemical potential $\mu_I=(\mu_u-\mu_d)/2$.

\begin{figure}[t]
	\centering
   \subfigure[]{%
	\label{fig:phaseDiagr}\includegraphics[width=0.475\textwidth]{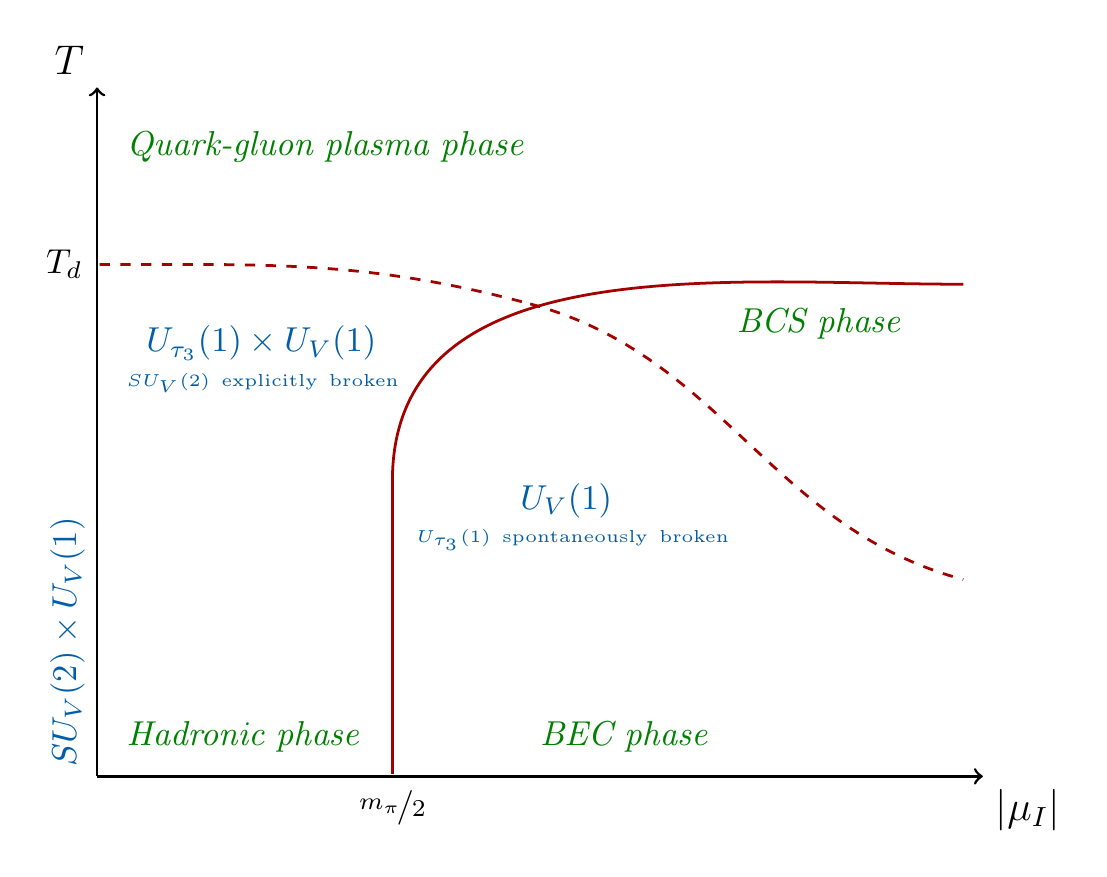}}
	\hfill
   \subfigure[]{%
	\label{fig:latticePhaseDiagr}\includegraphics[width=0.475\textwidth]{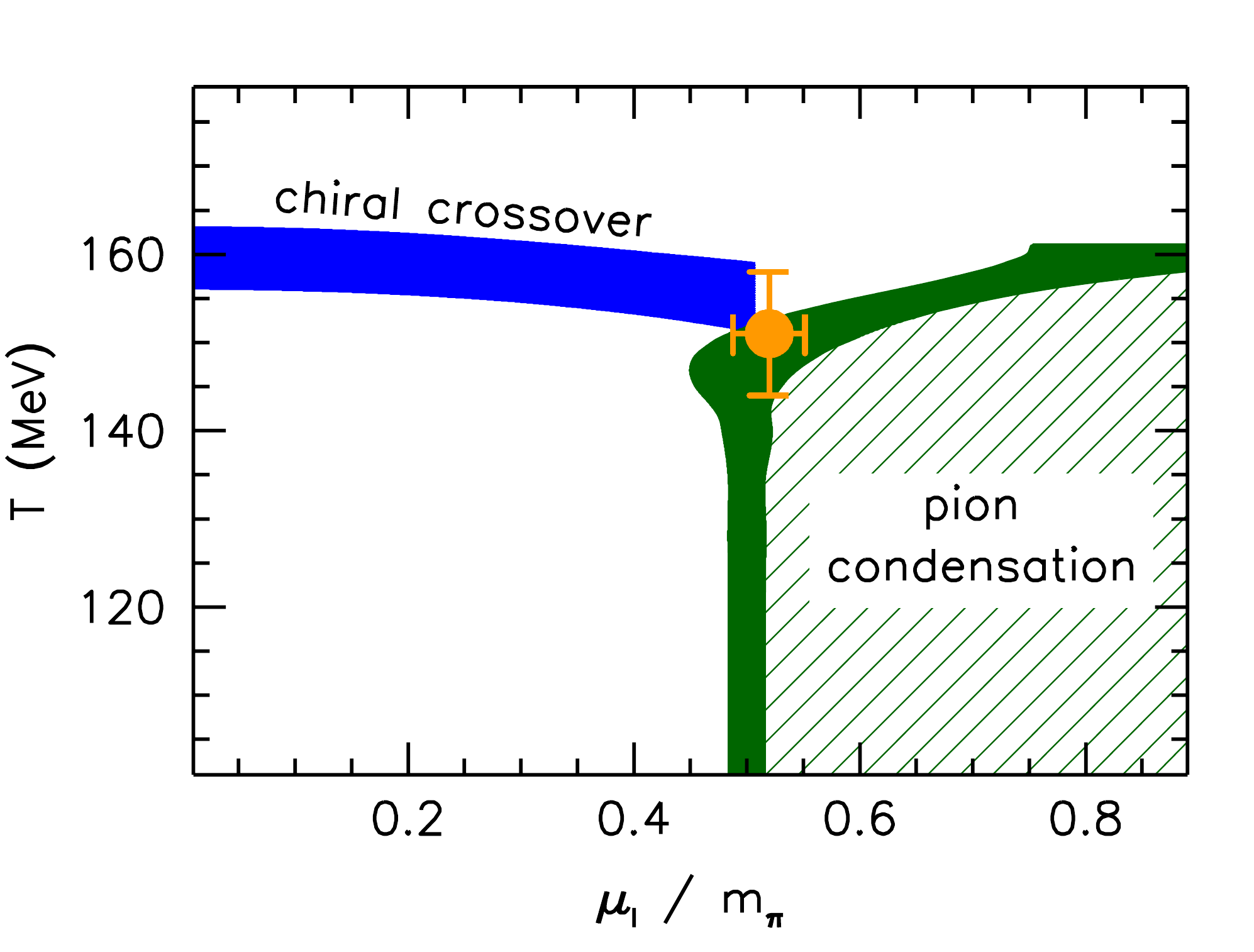}}
	\caption{Conjectured~\ref{fig:phaseDiagr}~\cite{Son:2000xc,Adhikari:2018cea} and measured~\ref{fig:latticePhaseDiagr}~\cite{Brandt:2017oyy,Brandt:2018omg} phase diagram of QCD at pure isospin chemical potential.}
\end{figure}

While systems of isospin-asymmetric matter are typically characterized by nonzero baryon density too, that is they also carry an excess of matter over antimatter encoded in a nonzero baryon chemical potential $\mu_B$, switching on $\mu_B$ would hinder direct lattice simulations due to the complex action problem.
As a first step it is then certainly useful to study the QCD phase diagram in the $(T, \mu_I)$ plane at $\mu_B=0$, which has the advantage of being fully accessible to standard lattice Monte Carlo techniques in principle. As anticipated by perturbation theory and model calculations~\cite{Son:2000xc,Adhikari:2018cea} (Fig.~\ref{fig:phaseDiagr}), lattice simulations found~\cite{Brandt:2017oyy,Brandt:2018omg} (Fig.~\ref{fig:latticePhaseDiagr}), an interesting and complex structure with at least three phases.
%, depending also on the temperature, from zero/small to intermediate/large $\mu_I$.

The main questions we address in this proceedings revolves around the existence of the BCS phase and the location of its boundaries.
The existence of a BCS phase is expected because perturbation theory, which is applicable in the limit $|\mu_I|\gg \Lambda_{QCD}$, predicts that the attractive gluon interaction forms pseudoscalar Cooper pairs of $u$ and $\bar d$ quarks at zero temperature~\cite{Son:2000xc}. 
Model calculations
%, e.g. a Polyakov-loop extended two-flavor quark-meson model,
also confirmed the existence of a BCS phase at nonzero temperature (see e.g.~\cite{Adhikari:2018cea}).
The transition between the BEC phase and the BCS phase is expected to be an analytic crossover, given that the symmetry breaking pattern is the same.
Lattice simulations also show large values for the Polyakov loop within the BEC phase~\cite{Brandt:2017oyy}. Those can be considered to be a hint for a superconducting ground state with deconfined quarks, that is for the BCS phase.
Here, we propose to look for further signatures for the existence and location of the BCS phase in the complex spectrum of the Dirac operator, which can be related to the BCS gap in a Banks-Casher  type relation~\cite{Kanazawa:2012zr} (cf. Eq.~\eqref{eq:BC}).

%%%%%%%%%%%%%%%%%%%%%%%%%%%%%%%%%%%%%%%%%%
\section{Simulation setup and observables}
The fermion matrix $\Ml$ within the action $S_{ud}=\bar\psi\Ml\,\psi$ for the light quarks $\psi=(u,d)^\top$ in Euclidean spacetime and in the continuum, reads
\begin{equation}
\Ml = \gamma_\mu (\partial_\mu + i A_\mu)\, \mathds{1} + \ml \mathds{1} + \mu_I \gamma_4 \tau_3  + i \lambda \gamma_5 \tau_2\,.
\label{eq:Sud}
\end{equation}
Here $A_\mu$ is the gluon field and $\tau_a$ are the Pauli matrices.
Note that besides the terms including the isospin chemical potential $\mu_I$ and the light quark mass $m_{ud}$, in $\Ml$ there also is an explicit symmetry breaking term including the parameter $\lambda$, referred to as pionic source, that couples to the charged pion field $\pion = \bar u \gamma_5 d - \bar d \gamma_5 u$.
This unphysical term is needed to enable the observation of the spontaneous breaking of the continuous $\mathrm{U}_{\tau_3}\!(1)$ symmetry and will act as a regulator for the simulations in the BEC phase~\cite{Kogut:2002tm,Kogut:2002zg,Endrodi:2014lja}.
Physical results are obtained by taking the $\lambda\to0$ limit.

For our measurements we consider 2+1-flavor QCD with $\mu_I>0$ and $\lambda>0$ as already simulated to map out the phase diagram shown in Fig~\ref{fig:latticePhaseDiagr}~\cite{Brandt:2017oyy}.
The lattices considered so far are $N_s^3\times N_t$ lattices with $N_t=6$ at various temperatures $T=1/(N_ta)$.
The Dirac operator is discretized employing the staggered formulation and the rooting procedure.
The partition function of this system is given in terms of the path integral over the gluon link variables $U_\mu=\exp(iaA_\mu)$,
\begin{equation}
\Z = \int \D U_\mu \, e^{-\beta S_G}\, (\det \Ml)^{1/4}\,(\det \Ms)^{1/4} \,,
\label{eq:Z}
\end{equation}
where $\beta=6/g^2$ is the inverse gauge coupling,  $S_G$ the tree-level Symanzik improved gluon action, $\Ml$ the light quark matrix in the basis of the up and down quarks and $\Ms$ the strange quark matrix,
\begin{equation}
\Ml
\!= \!
\begin{pmatrix}
 \Dp+\ml & \!\!\!\lambda \eta_5 \\
 -\lambda \eta_5 & \!\!\!\Dm +\ml
\end{pmatrix}, \qquad
\Ms \!=\! \Dn + m_s\,.
\label{eq:M}
\end{equation}
The argument of $\slashed{D}$ indicates the chemical potential $\mu_I$ and $\eta_5$ the staggered equivalent of $\gamma_5$.
The positivity of the integrand of $\Z$ can be shown. In particular, both determinants in the measure of the path integral~(\ref{eq:Z}) are positive. 
The quark masses are tuned to their physical values along the line of constant physics (LCP) from Ref.~\cite{Borsanyi:2010cj}, with the pion mass $m_\pi\approx 135\textmd{ MeV}$.

In the above setup our main observable is the spectrum of complex eigenvalues of the massless Dirac operator $\Dp$.
For the up quark, the eigenproblem reads
\begin{equation}
\Dp \,\psi_n = \nu_n\,\psi_n\,,
\label{eq:Dpp}
\end{equation}
where the eigenvalues $\nu_n$ are complex numbers.
The eigenproblem for the down quark can be obtained from Eq.~\eqref{eq:Dpp} using  chiral symmetry, i.e. $\Dp \eta_5 + \eta_5 \Dp = 0$, and hermiticity, i.e. $\eta_5 \Dp\eta_5 = \Dm^\dagger$ and reads
\begin{equation}
{\widetilde\psi_n}^\dagger \,\Dm = {\widetilde\psi_n}^\dagger\,\nu_n^*\,,\quad\quad
\widetilde\psi_n = \eta_5\psi_n\,.
\label{eq:Dmm}
\end{equation}
The fact that $[\Dp,\slashed{D}^\dagger(\mu_I) ] \neq0$, i.e. that $\Dp$ is not a normal operator, entails that its left and right eigenvectors do not coincide.
However, following Eqs.~(\ref{eq:Dpp}) and~(\ref{eq:Dmm}), for each eigenvalue in the up quark sector there is a complex conjugate pair in the down quark sector.

\begin{figure}[t]
	\centering
   \subfigure[]{%
	\label{fig:mui1}\includegraphics[width=0.325\textwidth]{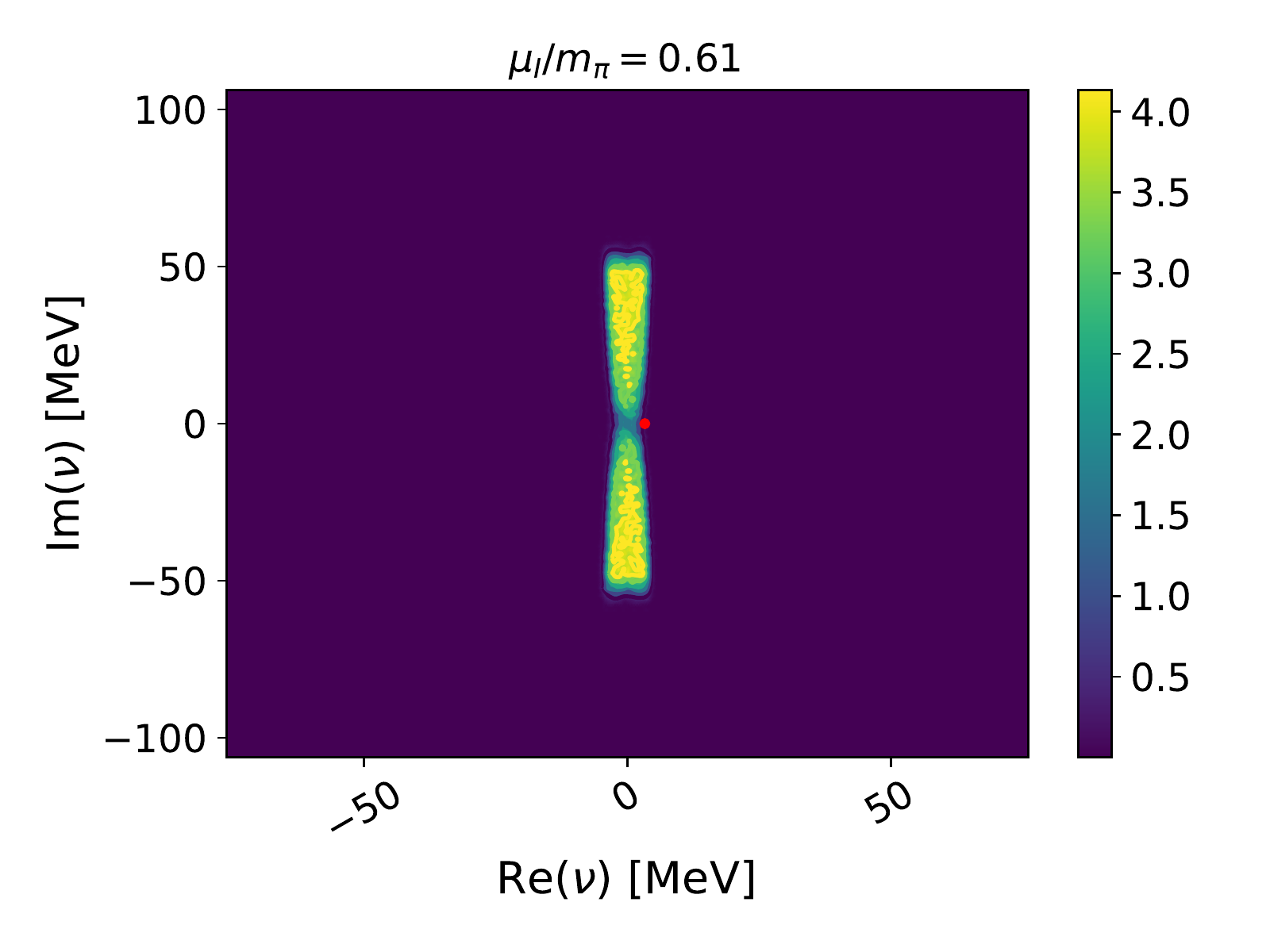}}
	\hfill
   \subfigure[]{%
	\label{fig:mui2}\includegraphics[width=0.325\textwidth]{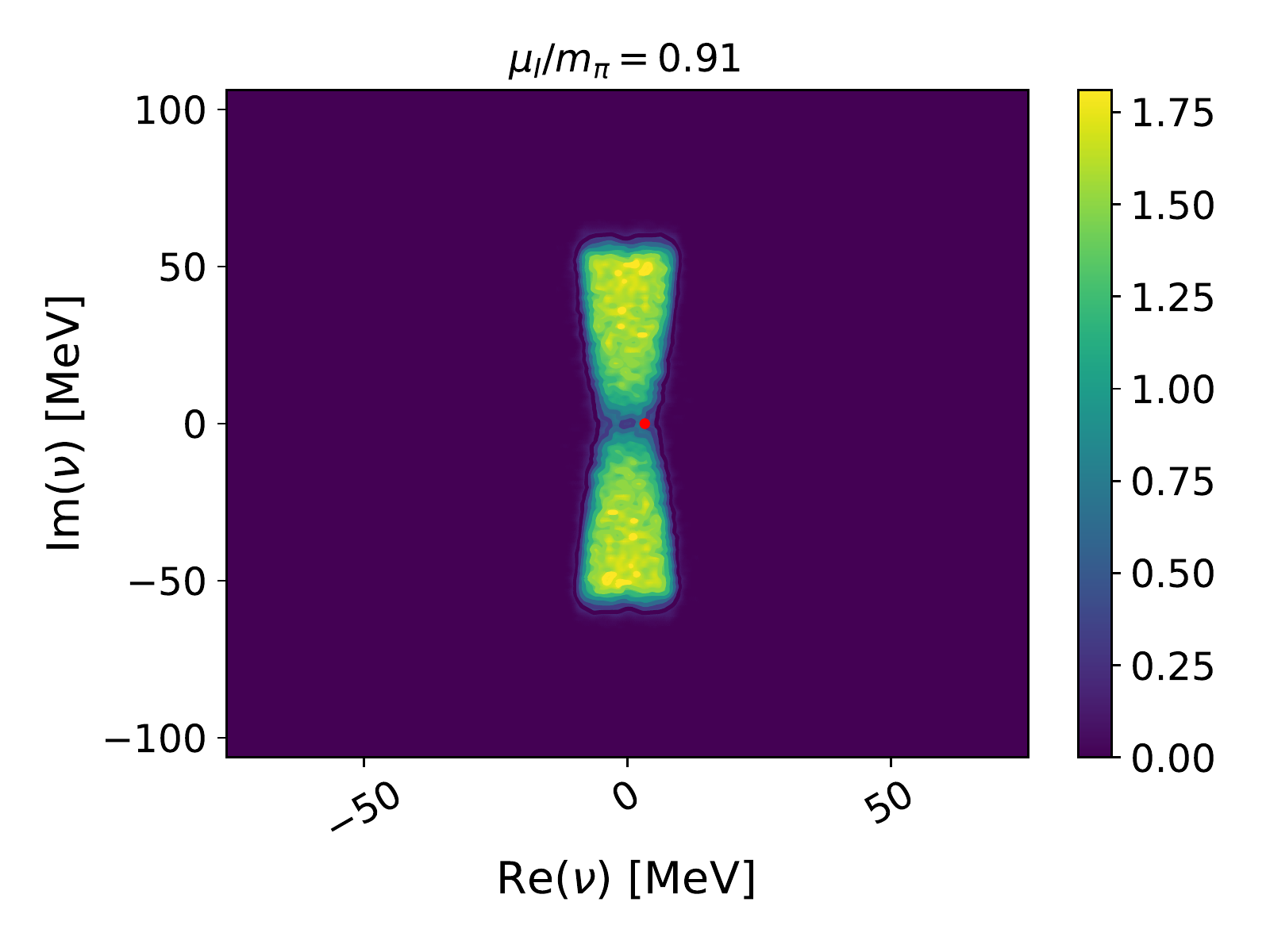}}
	\hfill
   \subfigure[]{%
	\label{fig:mui3}\includegraphics[width=0.325\textwidth]{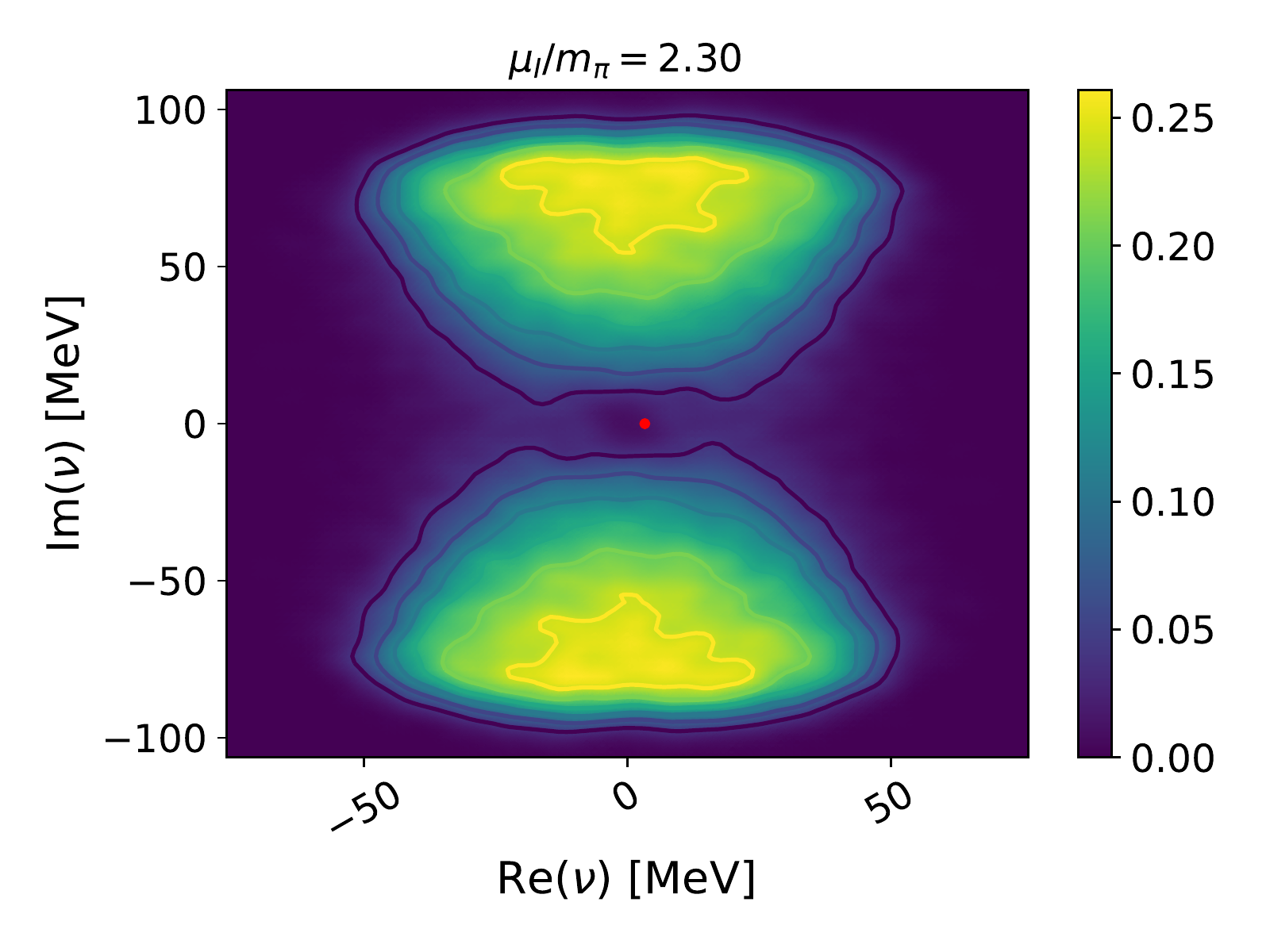}}
	\caption{\label{fig:mui}Contour plots of the complex spectrum of the Dirac operator as obtained for a $N_s=24$, $N_t=6$ lattice at $\lambda/\ml\sim0.29$ and $T=155$ MeV for isospin chemical potentials $\mu_I/m_\pi=0.61$~\ref{fig:mui1}, $\mu_I/m_\pi=0.91$~\ref{fig:mui2}, $\mu_I/m_\pi=2.30$~\ref{fig:mui3}. The red dot indicates $\ml + i \cdot 0$.}
\end{figure}

Our choice of observable is motivated by the extension of the Banks-Casher relation to the case of complex Dirac eigenvalues derived in Ref.~\cite{Kanazawa:2012zr} for the zero-temperature, high-density limits of QCD at nonzero isospin chemical potential.
The derived Banks-Casher type relation for massless quarks reads
\begin{equation}
    \Delta^2 = \frac{2\pi^3}{9}\rho(0).
    \label{eq:BC}
\end{equation}
This relation gives us a prescription on how to obtain information on the BCS gap $\Delta$ from the  density of the complex Dirac eigenvalues extrapolated at the origin $\rho(0)$.
In Ref.~\cite{Kanazawa:2012zr} the main idea for the extension of the Banks-Casher relation for $T=0$ and $|\mu_I|\gg \Lambda_{QCD}$ is to write down the partition function $\Z(M)$ as a function of the quark mass matrix $M$, both in the fundamental QCD-like theory and in the corresponding effective theory.
Taking suitable derivatives then yields an expression proportional to $\rho(0)$ in the fundamental theory and $\Delta^2$ in the effective theory.
The Banks-Casher-type relation is obtained by identifying these results which leads to Eq.~\eqref{eq:BC}. A similar relation is also expected to hold at nonzero quark masses and temperatures.

%%%%%%%%%%%%%%%%%%%%%%%%%%%%%%%%%%%%%%%%%%
\section{Results}

To solve the eigenproblem of Eq.~(\ref{eq:Dpp}) we employed the Scalable Library for Eigenvalue Problem Computations (SLEPc)~\cite{Hernandez:2005}, which is a software package for the solution of large sparse eigenproblems on parallel computers.
The solver used for the eigenvalue problem is a Krylov-Schur solver whose implementation within SLEPc is suited for non-Hermitian problems.
We compute about 150 eigenvalues of the non-hermitian Dirac operator, which are the closest (in modulo) to the origin.

\begin{figure}[t]
	\centering
   \subfigure[]{%
	\label{fig:density16}\includegraphics[width=0.475\textwidth]{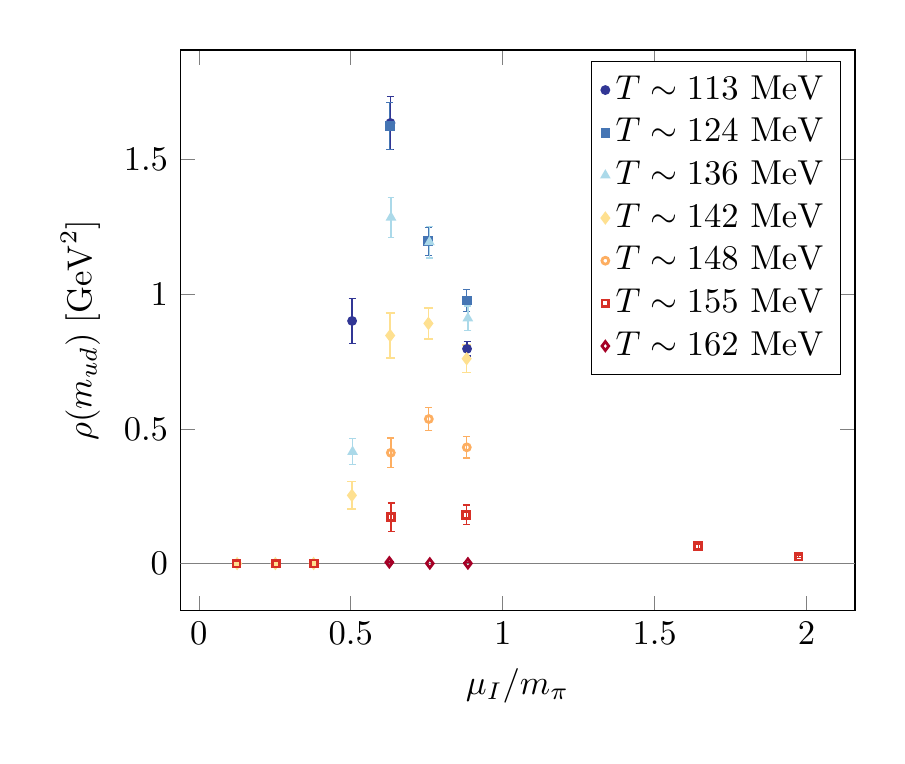}}
	\hfill
   \subfigure[]{%
	\label{fig:density24}\includegraphics[width=0.475\textwidth]{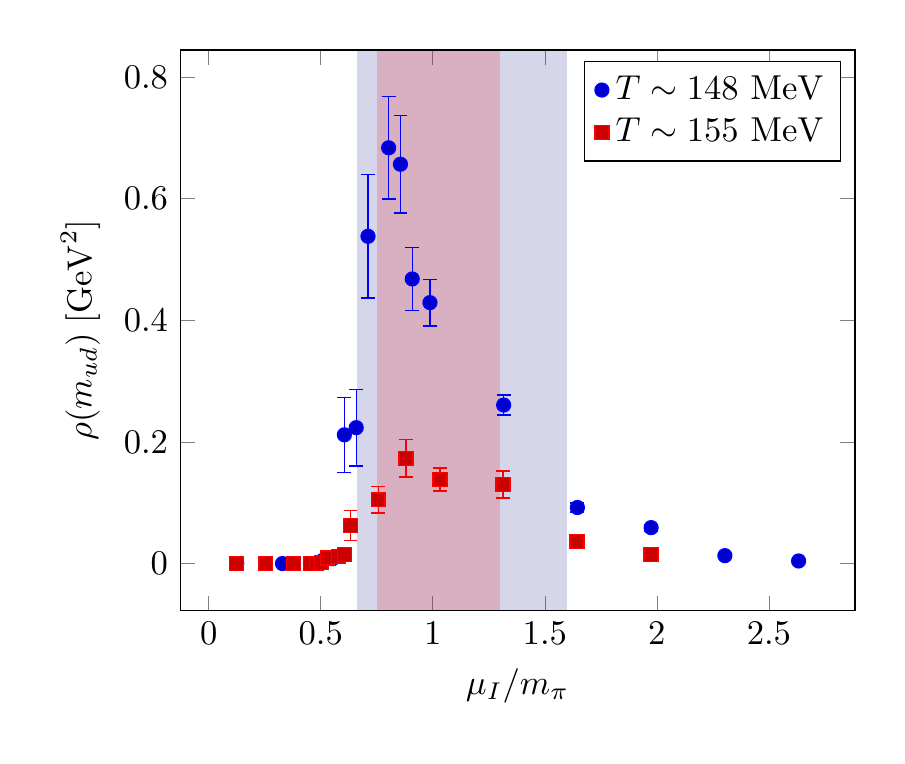}}
	\caption{\label{fig:density}$\rho(\ml)$ as a function of $\mu_I$, as obtained at various temperatures on $16^3\times6$~\ref{fig:density16} and $24^3\times6$~\ref{fig:density24} lattices. For the latter case only two temperature values are displayed. The lower(upper) edge of the shaded areas is set by the $\mu_I$ value at which the pion condensate $\Sigma_{\pi}$ becomes nonzero (the renormalized Polyakov loop $P_\mathrm{r}$ becomes 1) in the same setup.}
\end{figure}

Since the simulations are carried out for physical pion masses, away from the chiral limit (i.e. $\ml\neq0$), we try to extrapolate the density $\rho(\nu)$ to $\ml+i \cdot 0$ rather than to zero neglecting, at first, possible corrections due to non-zero masses and temperatures.
We evaluate $\rho(\ml)$ by using kernel density estimation (KDE), a non-parametric way to estimate the multivariate probability density function from the measured spectrum. Such technique is implemented in the python library scikit-learn~\cite{scikit}, which we employ for the analysis.

It can be observed, by inspecting the contour plots in Fig.~\ref{fig:mui}, how only for $\mu_I$ large enough, i.e. within the BEC phase, the spectrum is wide enough in the real direction to encompass the red dot in Fig.~\ref{fig:mui} at  $\ml$ resulting in $\rho(\ml)\neq0$.
At $\mu_I<m_\pi/2$ the eigenvalues are, instead, clustered along the imaginary axis and  $\rho(\ml)=0$.
At the largest simulated $\mu_I$ values there is a tendency $\rho(\ml)\to0$ due to the drift of the eigenvalues away from the real axis. However it should be noted that the impact of cutoff effects for larger and larger $\mu_I$ values remains to be assessed by a systematic comparison with results on finer lattices.

\begin{figure}[t]
	\centering
   \subfigure[]{%
	\label{fig:dataRhoPbGp}\includegraphics[width=0.475\textwidth]{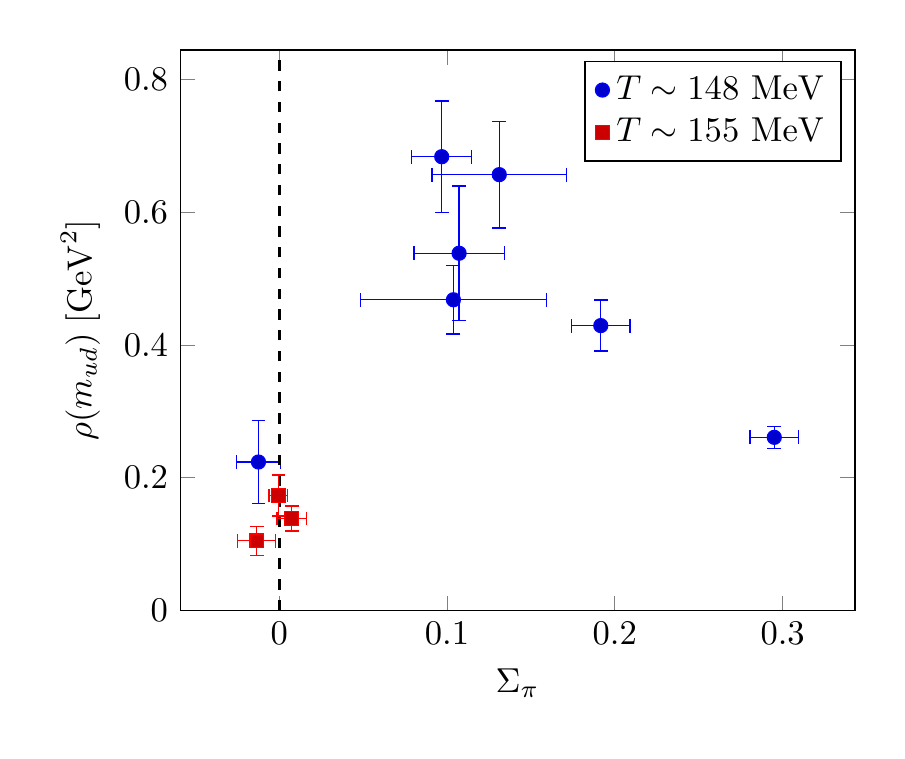}}
	\hfill
   \subfigure[]{%
	\label{fig:dataRhoPloop24}\includegraphics[width=0.475\textwidth]{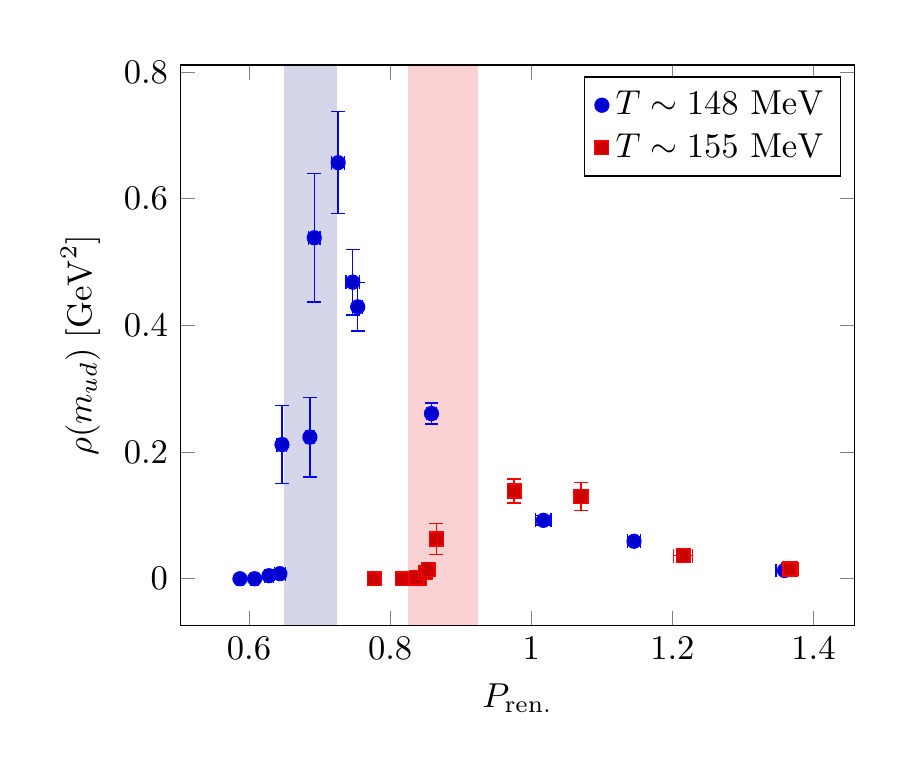}}
	\caption{\label{fig:dataRho}$\rho(\ml)$ as a function of $\Sigma_{\pi}$~\ref{fig:dataRhoPbGp} and of $P_{\mathrm{ren.}}$~\ref{fig:dataRhoPloop24}. For the latter case the shaded areas correspond to the range of values $P_{\mathrm{ren.}}$ takes within the green BEC boundary in Fig.~\ref{fig:latticePhaseDiagr} at the two considered temperatures.}
\end{figure}

Quantitative results for the spectral density are shown in Fig.~\ref{fig:density}.
It is interesting to match the $\mu_I$- and $T$- dependence of $\rho(\ml)$ with the location of the boundary of the BEC phase as determined by the onset of the pion condensate $\Sigma_{\pi}$ and with the location of the deconfinement crossover within the BEC phase as hinted for by a specific value of the renormalized Polyakov loop, that is $P_{\mathrm{ren.}}=1$. This is done both in Fig.~\ref{fig:density24} and in Fig.~\ref{fig:dataRho} by using results for $\Sigma_{\pi}$ and $P_{\mathrm{ren.}}$ obtained in the same setup in Ref~\cite{Brandt:2017oyy}.

\begin{figure}[t]
	\centering
   \subfigure[]{%
	\label{fig:dataRhoMuiVol}\includegraphics[width=0.475\textwidth]{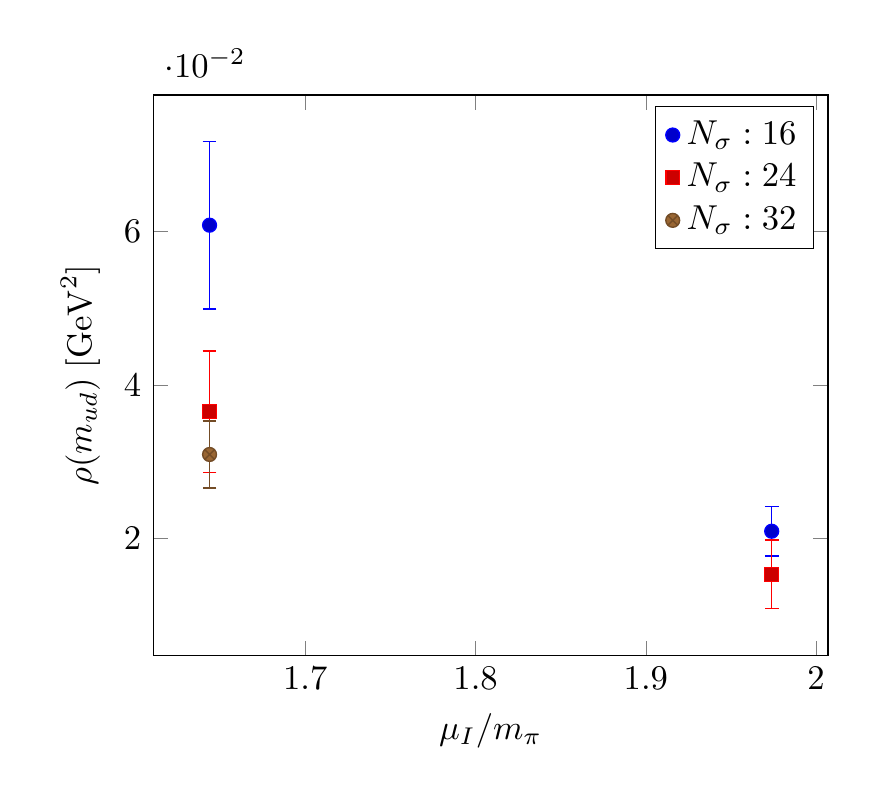}}
	\hfill
   \subfigure[]{%
	\label{fig:dataRhoMui24Lambda}\includegraphics[width=0.475\textwidth]{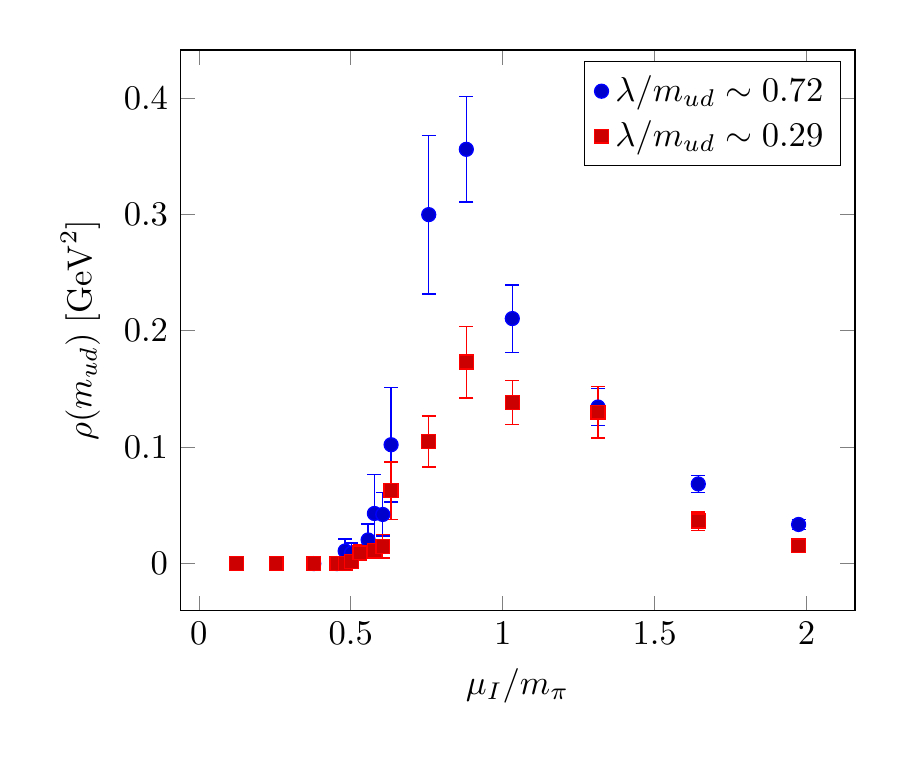}}
	\caption{\label{fig:dataRho2}\ref{fig:dataRhoMuiVol} $\rho(\ml)$ as a function of $\mu_I$ for $N_t=6$ and three different spatial volumes $N_s$. \ref{fig:dataRhoMui24Lambda} the $\lambda$ dependence of results for $T=155$MeV.}
\end{figure}

What can be observed is that the signal for the extrapolated spectral density seems to become nonzero around $\mu_I^{\mathrm{BEC}}(T)$, that is at the location of the BEC phase boundary for the considered temperature.
However, results also show that the extrapolated spectral density drops to zero again at larger values of $\mu_I$. Notice that lattice artefacts are expected to suppress $\rho(\ml)$, just as they do with $\Sigma_{\pi}$~\cite{Kogut:2002tm,Kogut:2002zg,Endrodi:2014lja}. Disentangling the signal for the BCS-BEC crossover from discretization errors at large $\mu_I$ is therefore difficult and a more systematic study is certainly needed to draw realistic conclusions.
As one can see from Fig.~\ref{fig:dataRho2}, the extrapolated spectral density still shows significant volume effects as well as a dependence of the results on the pionic source $\lambda$.

%%%%%%%%%%%%%%%%%%%%%%%%%%%%%%%%%%%%%%%%%%
\section{Discussion and conclusions}

The presented results clearly show that the extrapolated spectral density is sensitive to the BEC boundary. However, to be able to draw conclusions on whether, and in which $\mu_I$ range, there is sensitivity to the BEC-BCS crossover as well, a more systematic analysis is needed.
Such analysis will allow us to establish the expected quantitative connection between the measured density and the BCS gap.
Larger volumes, finer lattice spacings and a $\lambda\to0$ extrapolation must be considered, and this is ongoing work. Finer lattices, in particular, will help us identifying lattice artefacts due to cutoff effects at large $\mu_I$.
Moreover, given that the Banks-Casher relation that we intend to use as a prescription to connect the spectral density with the BCS gap is strictly valid only for $T=0$ and in the $|\mu_I|\gg \Lambda_{QCD}$ limit, a generalization of this relation away from this limit is desired. In addition, we might have to consider larger isospin chemical potentials and smaller temperatures.

%%%%%%%%%%%%%%%%%%%%%%%%%%%%%%%%%%%%%%%%%%
\vspace{6pt} 

%%%%%%%%%%%%%%%%%%%%%%%%%%%%%%%%%%%%%%%%%%
%% optional
%\supplementary{The following are available online at \linksupplementary{s1}, Figure S1: title, Table S1: title, Video S1: title.}


%%%%%%%%%%%%%%%%%%%%%%%%%%%%%%%%%%%%%%%%%%
%\authorcontributions{For research articles with several authors, a short paragraph specifying their individual contributions must be provided. The following statements should be used ``conceptualization, X.X. and Y.Y.; methodology, X.X.; software, X.X.; validation, X.X., Y.Y. and Z.Z.; formal analysis, X.X.; investigation, X.X.; resources, X.X.; data curation, X.X.; writing--original draft preparation, X.X.; writing--review and editing, X.X.; visualization, X.X.; supervision, X.X.; project administration, X.X.; funding acquisition, Y.Y.'', please turn to the  \href{http://img.mdpi.org/data/contributor-role-instruction.pdf}{CRediT taxonomy} for the term explanation. Authorship must be limited to those who have contributed substantially to the work reported.}

%%%%%%%%%%%%%%%%%%%%%%%%%%%%%%%%%%%%%%%%%%
\funding{The research has been funded by the DFG via the Emmy Noether Programme EN 1064/2-1.}

\reftitle{References}

\end{document}